\documentclass[runningheads]{llncs}

\usepackage{graphicx}
\usepackage{amsmath}
\usepackage{amsfonts}

\begin{document}
\title{Image-to-Graph Convolutional Network for Deformable Shape Reconstruction from a Single Projection Image}

\titlerunning{Image-to-Graph Convolutional Network from a Single Projection Image}
	
\author{Megumi Nakao\inst{1}\orcidID{0000-0002-5508-4366} \and
Fei Tong\inst{1} \and
Mitsuhiro Nakamura\inst{2}\orcidID{0000-0002-6406-2097} \and
Tetsuya Matsuda \inst{1}\orcidID{0000-0002-2339-1521}}

\authorrunning{M. Nakao et al.}

\institute{Graduate School of Informatics, Kyoto University, Kyoto, Japan.
\email{megumi@i.kyoto-u.ac.jp}\\
\and
Graduate School of Medicine, Human Health Sciences, Kyoto University, Kyoto, Japan.\\}

\maketitle

\begin{abstract}
Shape reconstruction of deformable organs from two-dimensional X-ray images is a key technology for image-guided intervention. In this paper, we propose an image-to-graph convolutional network (IGCN) for deformable shape reconstruction from a single-viewpoint projection image. The IGCN learns relationship between shape/deformation variability and the deep image features based on a deformation mapping scheme. In experiments targeted to the respiratory motion of abdominal organs, we confirmed the proposed framework with a regularized loss function can reconstruct liver shapes from a single digitally reconstructed radiograph with a mean distance error of $3.6 \text{~mm}$.

\keywords{Graph convolutional network, Shape reconstruction, Respiratory motion, X-ray image}
\end{abstract}

\section{Introduction}
Three-dimensional (3D) medical imaging is widely used for diagnosis and pre-treatment planning. However, organs can move or deform during surgery or radiotherapy, preventing accurate tumor localization and making precise treatment difficult. Treatment is performed using only two-dimensional (2D) images, such as endoscopic images or X-ray images, because of the limitations of available imaging devices. In this paper, we focus on 3D shape reconstruction from a single projection image, for image-guided interventions.

In the last decade, shape reconstruction for image-guided interventions has been investigated in two areas of research: 2D/3D registration for rigid bodies \cite{Miao16}\cite{Markelj12}\cite{Reyneke19} and deformable modeling of soft organs \cite{Koo17}\cite{Wu19}. In contrast to rigid registration, deformable registration has to handle point-to-point correspondence between  large numbers of voxels, making the application of deep learning difficult. For this issue, statistical modeling \cite{Reyneke19}\cite{Rigaud19}\cite{Nakamura21} based on curved manifolds or point-sampled mesh representations is a clinically important, computationally efficient approach for estimating the shapes of deformable organs. Wu et al. proposed a 3D shape reconstruction method based on a convolutional neural network (CNN) \cite{Wu19}, and showed that the 3D shape of the lungs during a pneumothorax deformation can be reconstructed from only a single-viewpoint image. However, because the shape was represented as point clouds, surface information  and topological information about the relationships  between vertices, which are important for deformation field computation, were lost.  Wang et al. proposed a CNN-based framework to calculate lung respiratory deformation from a digitally reconstructed radiograph (DRR) \cite {WangY20}. However, 3D shapes were artificially generated from multiple initial 3D templates with free-form deformation. Hence, the CNN-based reconstruction of organ shape for real patients has not yet been achieved or investigated except for our previous study \cite{Tong20}.

This paper introduces an image-to-graph convolutional network (IGCN) for deformable shape reconstruction from a single-viewpoint projection image. Specifically, the abdominal organs are hard to detect in low-contrast X-ray images, which contain considerable shape variability between patients. A 2D deformation mapping, an improved feature learning scheme is introduced to learn the relationship between 3D shape/deformation and the deep image features, while preserving the global shape using a regularized loss function. 

This study is the first to demonstrate the prediction performance of 2D/3D shape reconstruction targeted to the abdominal organs of real patients. We applied the IGCN to this task by using organ meshes, with point-to-point local correspondence obtained by deformable mesh registration (DMR) \cite{Nakao21}. deformation. Because the IGCN can generate the shape of a 3D organ in real time, for example, in radiotherapy, it can be used to estimate the area of organs at risk from only X-ray images or perform tumor localization despite respiratory 

\section{Methods}

\subsection{Dataset and Problem Definition} 
\begin{figure*}[t]
    \centering
    \includegraphics[width=12.0cm]{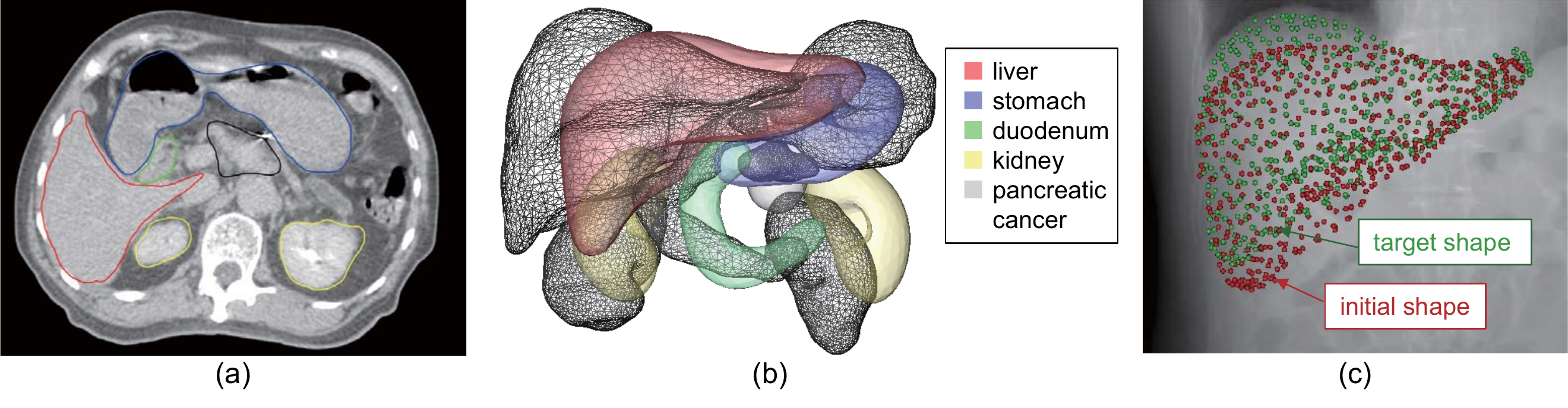}
    \caption{Abdominal CT dataset and preprocessing. (a) CT slice image with 3D contours of the liver, stomach, duodenum, and kidneys. (b) Statistically reconstructed mean shapes (translucent) and patient-specific models (mesh). (c) DRR of the target state combined with the projected vertices of the initial and target state.}
    \label{fig:1}
\end{figure*}

3D-CT volumes of 124 cases and 4D-CT volumes of 35 cases, acquired from different patients who underwent intensity-modulated radiotherapy in Kyoto University Hospital, were collected. Each 4D-CT volume used in this study consists of two time phases (the end-inhalation and end-exhalation phases) of 3D-CT volumes. Each 3D-CT volume consists of $512 \times 512$ pixels and 88–152 slices (voxel resolution: $1.0 \text{~mm} \times 1.0 \text{~mm} \times 2.5 \text{~mm}$). During routine clinical procedures, the regions of the entire body, stomach, liver, duodenum, left and right kidneys, and the clinical target volume (CTV) were labeled by board-certified radiation oncologists, as shown in Fig. \ref{fig:1}(a). 

We generated the surface meshes (400--500 vertices and 796--996 triangles for an organ) from the region labels, and obtained organ mesh models with point-to-point correspondence using DMR. The DMR algorithm and the registration performance for the abdominal organ shapes were published in a previous paper \cite{Nakao21}, and it was confirmed that a template mesh was registered to patient-specific organ shapes with  a $0.2 \text{mm}$ mean distance error, and $1.1 \text{mm}$ Hausdorff distance error, on average. Because the obtained models have point-to-point correspondence, the average shape can be obtained by calculating the average of each coordinate. Fig. \ref{fig:1}(b) shows one example of the registered organ models of one patient (mesh) and mean shapes (translucent) computed from all registered models. 

Fig. \ref{fig:1}(c) is a part of DRR with $640\times640$ pixels generated from a 3D-CT volume of the target state, overlaid with the projected vertices of the initial and target shape. The patient's body is fixed to the reference position so that the CTV in the initial state corresponds to the focus of radiation beams. Therefore, we assume that the camera parameters (i.e., the projection matrix) for generating the DRR are given, and the organ models can be projected to the DRR using the CTV center in the end-inhalation state as the origin. The shape of the diaphragm visualized in the DRR does not match the projected initial shape because the two states are the most distant in terms of the respiratory phase. The respiratory motion contains nonlinear deformation with local rotation and sliding motion \cite{Nakao21}\cite{Jud17}, and simple linear transformation is not sufficient to register these two states.

The IGCN is designed as a generalized, organ-independent framework. Although the reconstruction performance for all organs could be investigated, we first targeted liver shapes with partly detectable features (i.e., the diaphragm), to discuss how reconstruction error occurs locally in the cases with and without visual cues. The shape and location of the liver in the end-exhalation phase is the prediction target, and the shape in the end-inhalation phase is the initial patient anatomy used as the input data. This problem definition is important for investigating the prediction performance in 2D/3D deformable organ reconstruction because nonlinear deformation between the two states \cite{Nakao21} can cause the maximum prediction error in one respiratory cycle.

Because 4D-CT volumes of 35 cases are not sufficient for learning the relationship between shape/deformation variability and the 2D projection images, we created an augmented training dataset from the 3D-CT images of another 124 cases. The 4D-CT volume dataset was divided into a training set (20 cases) and test set (15 cases). The mean and standard deviation of the vertex displacement were obtained from the training data, and a similar global translation, generated using Gaussian noise, was applied to all the 3D-CT liver models. A total of 144 datasets (20 4D-CT volumes and 124 volumes from the augmented 3D-CT data) were used to train the IGCN network.

\subsection{Image-to-Graph Convolutional Network}

\begin{figure}[t]
    \centering
    \includegraphics[width=12.0cm]{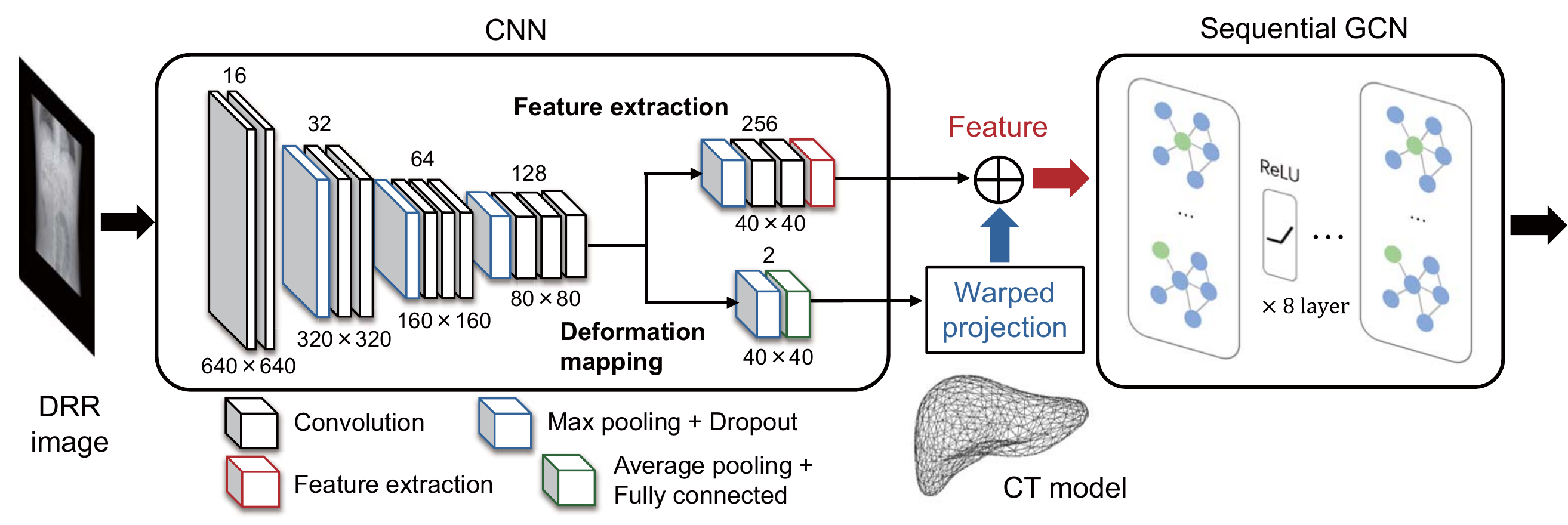}
    \caption{The full image-to-graph convolutional network. The CNN is used for extracting image features from a single projection image, and the GCN is for learning 3D deformation. $\bigoplus$ represents the concatenation of image features and vertex coordinates.}
\label{fig:2}
\end{figure}

Fig. \ref{fig:2} shows the IGCN, which consists of a CNN that extracts perceptual features from the input image and a graph convolutional network (GCN) that learns mesh deformation according to the extracted image features. In this framework, the initial model is projected onto the input DRR image. The image features corresponding to each vertex ${v}_{i}$ can be extracted. The image features and vertex coordinates are concatenated and incorporated into the GCN for learning deformation. Both networks are optimized simultaneously using a loss function. We used an extended VGG-16 model\cite{Simonyan14} without pretraining. The width (number of channels) of each CNN layer is marked above the layers, and the size of each layer is marked below. 

Pixel2Mesh (P2M) employs a similar network architecture with hierarchical extension to fit an ellipsoid template to a variety of 3D objects \cite{P2M}. However, it concentrates on mesh deformation and does not consider movement of the target object, meaning that image features distant from the initial template are not learned. This is inadequate for our application because our prediction target contains both local deformation and global translation. In addition, the X-ray images have no clear edges in most parts of the organs. To address these complications, we introduce an improved feature learning scheme using 2D deformation mapping and use a regularized loss function to predict respiratory-associated deformation accurately. 

\begin{figure}[t]
      \centering
      \includegraphics[width=12.0cm]{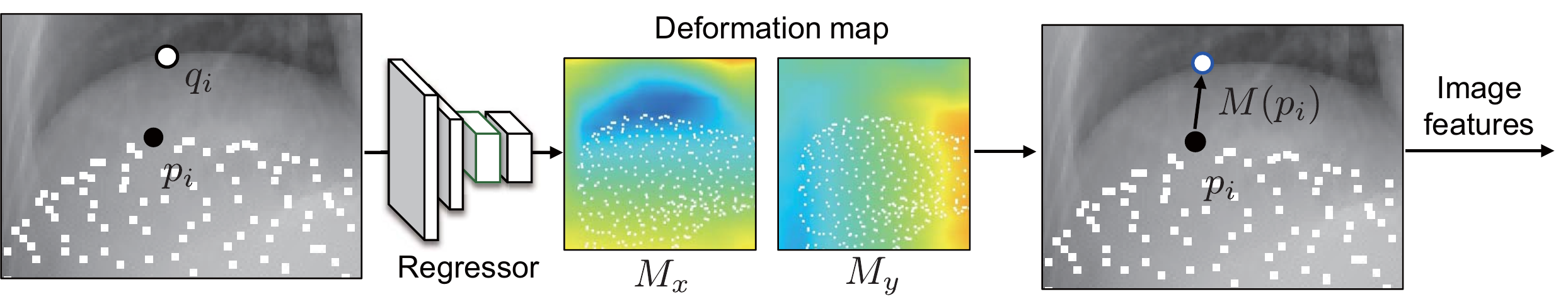}
      \caption{The image feature learning scheme using 2D deformation mapping for distant target. A part of CNN layer is used to estimate the deformation map over all vertices.}
      \label{fig:3}
\end{figure}

Because the patient's initial liver shape $M_I$ is determined from 3D-CT images in our framework, $M_I$ is used as the initial template for the IGCN. With the projection matrix, reflecting the camera parameters, each vertex ${v}_{i}$ is projected to the corresponding 2D pixel coordinate $p_{i}$ in the DRR image space (see Fig. \ref{fig:3}). P2M uses the projected point $p_{i}$ of the initial mesh template to capture image features; however, the corresponding features are distant from $p_{i}$ because of the displacement of the target organ. P2M relies on convolution to capture distant image features but convolution is not effective for capturing high-resolution features at the ideal, corresponding position $q_i$. 

We introduce a deformation mapping, a new concept to overcome the above limitation. The projection point is mapped to a new position $M(p_i)$ (called warped projection in Fig. \ref{fig:2}), where a higher probability of obtaining effective image features is expected. The deformation map is a learnable spatial mapping function determined by the 2D vector field $(M_x, M_y)$. The color map in Fig. 3 represents the mapping function $M$, describing the learned displacement in the $x$ and $y$ directions in 2D-projection image coordinates. This scheme was implemented as an extension to the feature extraction scheme of the CNN part.

For the GCN layers that generate the predicted 3D organ shapes, graph convolution is applied to obtain hierarchical topological features in non-Euclidean space \cite{Kipf17}. The mesh is a type of graph ${G(\mathcal{V}, \mathcal{E})}$, where $\mathcal{V}$ is the set of vertices and $\mathcal{E}$ is the set of edges. Per-vertex features are shared with neighbor vertices. The GCN in our study consists of eight sequential graph convolutional layers, each of which is defined in Eq. (1).
\begin{eqnarray}
    X^{(l+1)}=\sigma(\hat D^-{}^\frac{1}{2}\hat A \hat D^-{}^\frac{1}{2} X^{(l)} W^{(l)}) ,
\end{eqnarray}   
where $X^{(l)}$ and $X^{(l+1)}$ denote the feature matrix before and after convolution. ${A}\in \mathbb{R}^{n \times n}$ (where $n$ is the number of vertices) is the adjacency matrix: a symmetric matrix with binary values, in which element $A_{ij}$ is 1 if there is an edge between $v_i$ and $v_j$, or 0 if the two vertices are not connected. ${D}\in \mathbb{R}^{n \times n}$ is the degree matrix: a diagonal matrix, in which each element $A_{ii}$ represents the number of edges connected to $v_i$. $W$ is the learnable parameter matrix and the feature $X^{(l)}$ is the concatenation of 2D image features from the CNN and 3D vertex coordinates. The initial shape is deformed by updating $X^{(l)}$.

\subsection{Loss Functions}
We introduce three loss functions to constrain mesh deformation and projection point registration. Unlike the P2M framework \cite{P2M}, which is intended for generating target surfaces without corresponding vertices, in our framework, the ground-truth positions of the target models are obtained from the deformable registration process. For a strict evaluation of point-to-point correspondence, we define the mean distance loss $\mathcal{L}_{pos}$ of vertex positions between the estimated shape and the ground truth. $\mathcal{L}_{pos}$ is defined by
\begin{eqnarray}
       \mathcal{L}_{pos}=\frac{1}{n} \sum_{i=1}^{n}\|v_{i}-\hat{v}_{i}\|^{2}_{2} ,
\end{eqnarray}
where $n$ is the number of vertices, 
$v_{i} \in \mathcal{V} (i = 1,2,...,n)$ is the target 3D position, and $\hat{v}_{i}$ is the predicted position. This loss function induces the convergence of the estimated vertex to the correct position.

In addition to evaluating the similarity of 3D surfaces, we found that matching the projected 2D points improves feature learning and accuracy of 2D/3D reconstruction results. Specifically, stable learning of the deformation map is important when the target contains both translation and local deformation. We introduce the mapping loss $\mathcal{L}_{map}$, based on the projected points.

\begin{eqnarray}
       \mathcal{L}_{map}=\frac{1}{n} \sum_{i=1}^{n}\|q_{i} - M(p_{i})\|^{2}_{2} ,
\end{eqnarray}
where $p_{i}$ is the projected point of the initial shape, and $q_{i}$ is the projected point that corresponds to the target vertex ${v}_{i}$. $M$ is the mapping function trained by the CNN. The organ deformation is expected to remain within a limited range in our problem setting. To preserve the curvature and smoothness of the initial surface, we use a regularization term that evaluates a discrete Laplacian of the mesh. The Laplacian loss $\mathcal{L}_{laplacian}$ is defined as follows.
 \begin{eqnarray}
       \mathcal{L}_{laplacian} = \frac{1}{n}\sum_{i=0}^{n}\|L(v_{i})-L(\hat{v}_{i})\|^{2}_{2} ,
 \end{eqnarray}
where $L (\cdot)$ is the Laplace–Beltrami operator and $L(v_i)$ is the discrete Laplacian of the vertex $v_{i}$ defined by $L(v_i) = \sum_{j \in N(v_i)} (v_i -v_j)/N(v_i)$. $N(v_i)$ is the number of adjacent vertices $v_j$ of the 1-ring connected by the vertex $v_i$. This loss constrains the shape changes from the initial state and avoids the generation of unexpected surface noise and low-quality meshes.

The values of loss functions are normalized using the maximum values in each coordinate. The total loss is the weighted sum of three loss functions:
\begin{eqnarray}
    \mathcal{L}_{total} = \mathcal{L}_{pos}+\lambda_{map}\mathcal{L}_{map}+\lambda_{laplacian}\mathcal{L}_{laplacian}.
\end{eqnarray}
To facilitate feature learning using deformation mapping, we used 10.0 for $\lambda_{map}$ and 1.0 for $\lambda_{laplacian}$ after examination of several parameter sets.

\section{Experiments}
In the experiments, the performance of 2D/3D shape reconstruction of the liver was confirmed while comparing it with the results from the existing end-to-end deep learning framework. The whole network was implemented with Python 3.6.8, TFLearn, and the TensorFlow GPU library. The network was trained using an Adam optimizer with a learning rate of $1\times 10^{-4}$. The batch size was 1, and the total number of training epochs was 1000. 0.5 was used for the dropout rate. The training took 4.5 hours on a single NVIDIA GeForce RTX 2070. 

The implemented IGCN framework can provide the 3D meshes from the initial shape and the DRR in the target state. In this study, the mean distance (MD) between surfaces \cite{Rigaud19}, the root mean square error (RMSE) between corresponding vertices, and the Dice similarity coefficient (DSC) were used as the shape similarity metrics. MD is the mean value of the shortest bidirectional point-to-surface distance, and DSC measures the volume overlap between the deformed meshes and the ground-truth meshes. We compared the performance of the proposed IGCN framework, with and without 2D deformation mapping, and P2M \cite{P2M}. It should be noted that the correct position of each vertex was obtained from registered models. Hence, in this comparison, we replaced the Chamfer loss (used by P2M) with the mean distance loss defined by Eq. (2), leaving the remaining losses in P2M unchanged. Hierarchical learning was not used to match the end-to-end prediction process of the methods.

\begin{table}[t]
    \begin{center}
    \caption{Quantitative comparison of the shape reconstruction performance. The mean$\pm$standard deviation of MD, RSME, and DSC for predicted and target shapes. }
\scalebox{0.9}{    
    \begin{tabular}{ccccccccc}
    \hline
    & & Initial & & P2M & & IGCN (no mapping) & & IGCN \\ \hline
        
    MD [mm] & & 5.7 $\pm$ 2.9 & & 5.1 $\pm$ 1.5 & & 3.9 $\pm$ 0.7 & & 3.6 $\pm$ 1.2  \\
    RMSE [mm] & & 12.1 $\pm$ 5.3 & & 9.9 $\pm$ 5.1 & & 9.3 $\pm$ 1.8 & & 8.4 $\pm$ 2.2  \\
    DSC [\%] & & 84.2 $\pm$ 7.6 & & 86.8 $\pm$ 4.0 & & 89.4 $\pm$ 2.0 & & 91.5 $\pm$ 3.4 \\ \hline
    \end{tabular}
}
  \label{table:1}
  \end{center}
\end{table}
\begin{figure}[t]
      \centering
      \includegraphics[width=12.0cm]{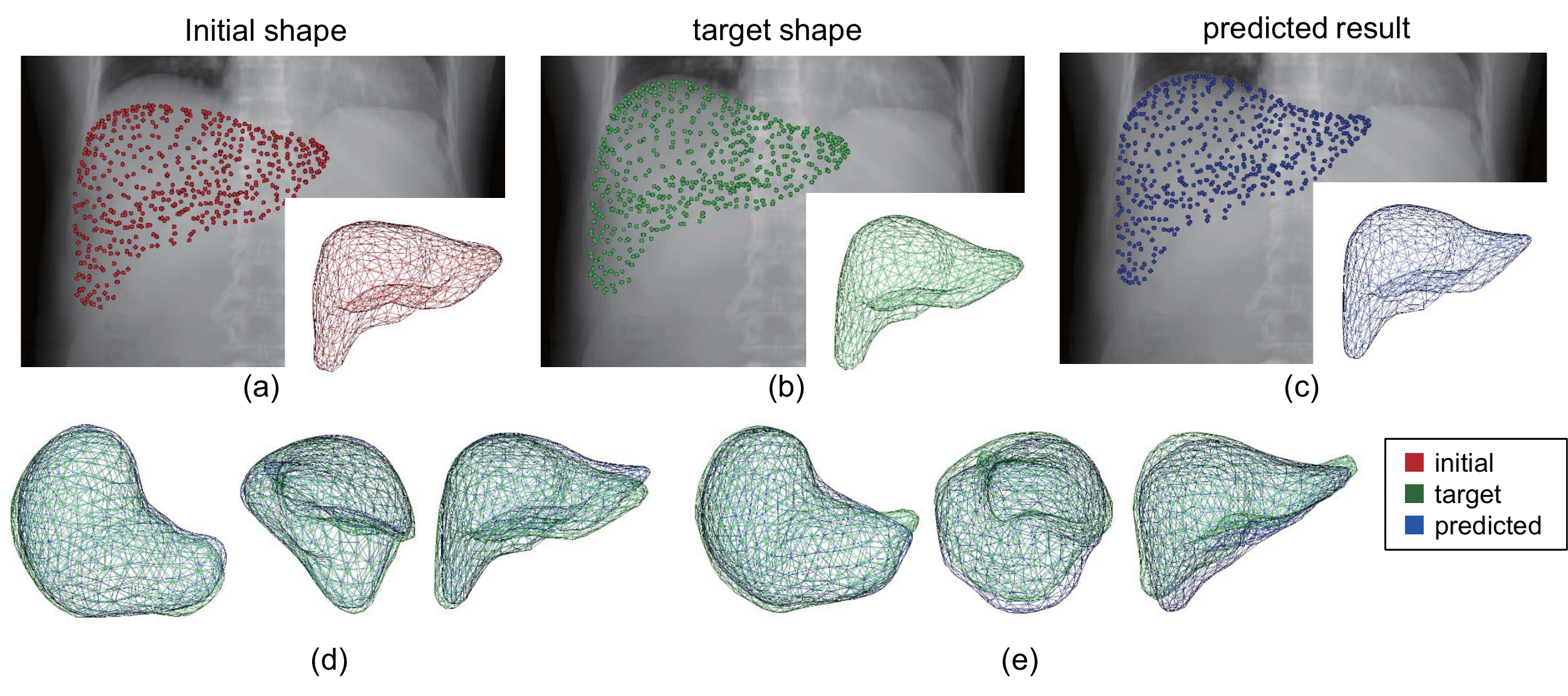}
      \caption{3D shape reconstruction results. (a) Initial liver shape and its projection to target DRR image. (b) Target shape. (c) Predicted liver shape with the average deformation error. (d) Deformation error between target and estimation result. (e) Estimation result with the largest deformation error.}
      \label{fig:4}
\end{figure}

\begin{figure}[t]
      \centering
      \includegraphics[width=11.5cm]{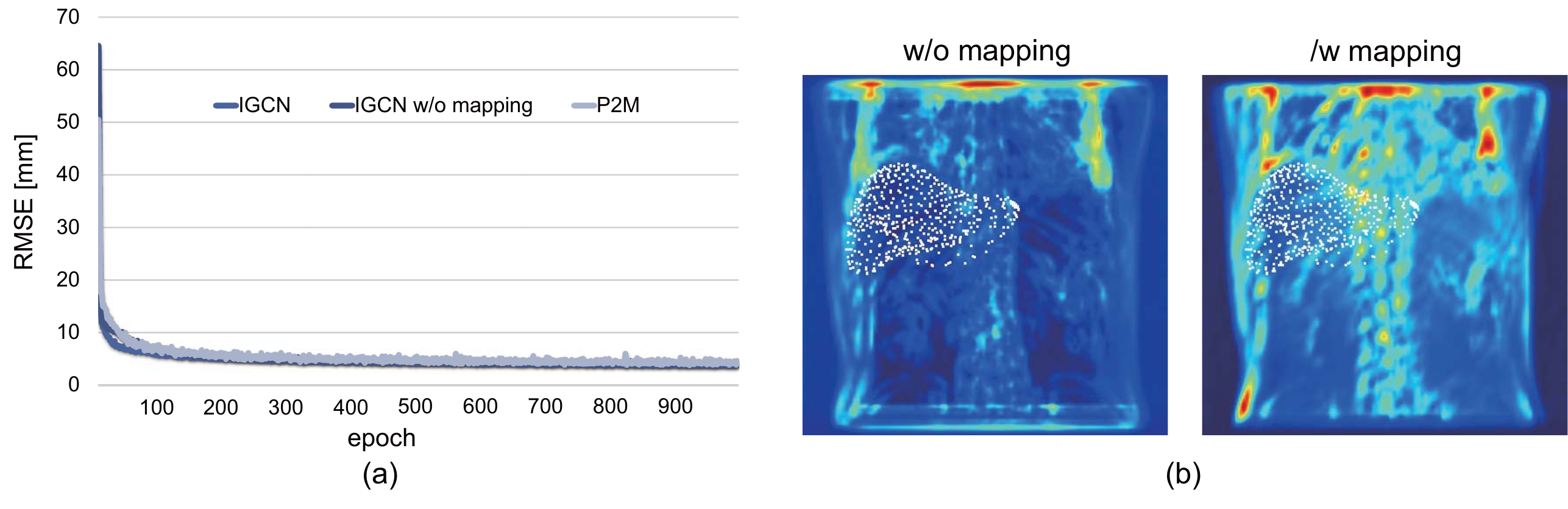}
      \caption{Training results. (a) Learning curves showing RMSE trained in 1000 iterations. (b) Averaged output features of the 128-channel layer learned with and without deformation mapping.The projected points of the initial shape are overlaid for reference.}
      \label{fig:5}
\end{figure}

The evaluation results for the 15 test cases are listed in Table \ref{table:1}. The results show that MD, RMSE, and DSC for the IGCN framework are significantly smaller than those obtained with the P2M (oneway analysis of variance, ANOVA; $p < 0.05$ significance level). Because these two metrics do not reflect the smoothness or estimated shape quality, we visualize the estimated shape in Fig.  \ref{fig:4}, to better assess these aspects of performance. The 3D shape was projected onto the DRR, with the pixels corresponding to each 3D vertex colored in red, green, and blue, corresponding to initial, target, and predicted, respectively. Although the projection of the initial shape contains considerable errors for radiotherapy, the estimated 3D shape has been improved, particularly the diaphragm part, where an obvious contour can be observed. The overlap between the predicted shape and the target shape is shown in Fig. \ref{fig:4}(d) and (e). Despite that only very low-contrast textures are confirmed in most parts of the liver, the deformation can be spatially reconstructed. The graph convolutions embed per-vertex features with connected neighbors, which results in better estimation performance of the regions with no visual cues. We measured the computation time for the whole shape reconstruction process performed in the CNN and GCN layers. The average computation time was 35.4 ms (28 frames per second), demonstrating the real-time performance of the IGCN.

Fig. \ref{fig:5} (a) demonstrates the training curves of the three methods for RMSE of training datasets. Each model closely followed and converged before 1000 iterations, with IGCN converging fastest and P2M showing a slightly unstable curve. On the other hand, the deep features learned from DRR are different, and it was confirmed that low-contrast textures such as the spine and vessel structures were extracted in addition to the diaphragm, as shown in Fig. 5 (b). We tried other conditions and settings, but the performance of P2M was not improved. This is because P2M does not use dropout, and there is a possibility of overfitting by the number of training data against the number of layers of CNN.

\section{Conclusion}
This paper proposed IGCN that combines a GCN with a CNN, to reconstruct the 3D shape of organs from low-contrast, 2D projection images. To achieve stable and accurate shape reconstruction, we introduced an improved feature learning scheme using deformation mapping and a newly designed loss function. Our future work includes the performance analysis of our method on other abdominal organs.

\subsection*{Acknowledgments}
This research was supported by a JSPS Grant-in-Aid for Scientific Research (B) (Grant number 18H02766a and 19H04484). We thank Edanz Group (https://en-author-services.edanz.com/ac) for editing a draft of this manuscript.

\end{document}